Magnetic penetration depth in electron-doped cuprates - evidence for gap nodes

Ruslan Prozorov[a], Russell W. Giannetta[a], Patrick Fournier[b], and Richard L. Greene[b].

[a]Loomis Laboratory of Physics, University of Illinois at Urbana-Champaign,
1110 West Green St, Urbana, Illinois 61801.

[b]Center for Superconductivity Research, Department of Physics, University of Maryland,
College Park, Maryland 20742.

The in-plane penetration depth $\lambda(T)$ is measured in electron-doped single crystals $Nd_{1.85}Ce_{0.15}CuO_{4-x}$ (NCCO) and $Pr_{1.85}Ce_{0.15}CuO_{4-x}$ (PCCO) using a 11 *MHz* LC resonator. In NCCO, $\lambda(T)$ exhibits a minimum at 3.8 *K* and a pronounced upturn down to 0.4 *K* due to the paramagnetic contribution of $Nd^{3+}$ ions. The London penetration depth contribution is linear in *T*. The paramagnetic contribution is absent in PCCO, where $\lambda(T) \sim T^2$ at low temperatures. Our results indicate the presence of nodes in the superconducting gap, i.e., non *s*-wave symmetry of the order parameter in electron-doped cuprates.

1.  INTRODUCTION

Although, electron doped cuprates [1] are believed to be s-wave superconductors [2-4], a literature review reveals that this issue remains controversial. Many tunneling measurements support an *s*-wave picture [3] while other STM data shows a zero bias conductance peak and support for *d*-wave pairing [5]. The London penetration depth $\lambda_L(T)$ gives a precise measure of the quasiparticle population. Previous $\lambda_L(T)$ measurements in NCCO appeared to reveal an exponential suppression of quasiparticles at low temperatures, supporting an s-wave picture [2]. However, Cooper pointed out that in NCCO, the paramagnetism of $Nd^{3+}$ ions could mask the power law dependence of $\lambda_L(T)$ expected for a *d*-wave superconductor [6]. Recent measurements of the penetration depth in NCCO films, using grain boundary junctions, have clearly shown the paramagnetic effect predicted by Cooper [4]. However, the authors claim that the resulting $\lambda_L(T)$, after correcting for paramagnetism, is still exponential and therefore both NCCO and PCCO are *s*-wave superconductors.

In this paper, we report high precision measurements of $\lambda_L(T)$ in single crystals of both NCCO and PCCO down to 0.4 *K*. Our data reveal power law dependences for $\lambda_L(T)$ in both materials and thus strong evidence for nodes in the energy gap.

2.  EXPERIMENTAL

Measurements were performed using a $f_0$=11 *MHz* LC resonator operating in a $^3$He refrigerator [7,8]. Detailed description of the apparatus is given in Ref. [8]. The overall noise level $\Delta f/f_0 \approx 10^{-9}$ enables to resolve $\Delta\lambda < 0.1$ Å for typical samples. The relative resonance frequency shift $\Delta f = f(T) - f(T_{min})$ is related to the change of the magnetic penetration depth via $\Delta\lambda = -G\Delta f$, where *G* is the sample and apparatus dependent calibration constant [8].

## 3. RESULTS

Low temperature behavior of the penetration depth for PCCO single crystal is shown in Fig.1. The inset shows the full range variation above the transition temperature. The solid line is a fit to $\Delta\lambda(T) \sim T^2$ - indistinguishable from the data up to $T/T_c=0.32$ above which the low-temperature approximation is not valid.

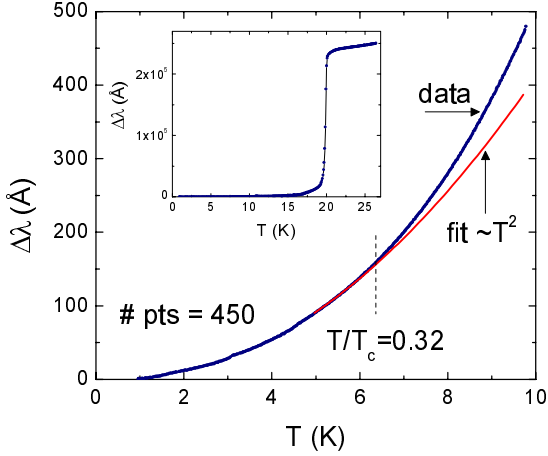

Figure 1. Variation of the penetration depth in PCCO single crystal. *Inset:* Full temperature range.

The quadratic temperature dependence of $\Delta\lambda(T)$ is consistent with *d*-wave superconductor in the presence of unitary limit impurity scattering [9].

Figure 2 shows $\Delta\lambda(T)$ for single crystal NCCO. Below $T=3.8$ K there is a pronounced upturn, which we attribute to the paramagnetic contribution of $Nd^{3+}$ ions [6]. According to Cooper [6], if the magnetic permeability $\mu(T)=1+4\pi\chi(T)$ is significant, the measured penetration depth is given by $\lambda(T)=\lambda_L(T)\sqrt{\mu(T)}$ [6]. Solid lines in Fig.2 depict fits to different models taking into account the $\sqrt{\mu(T)}$ term. Clearly, the standard *s*-wave isotropic BCS curve with $\varepsilon=2\Delta/T_c=3.53$ does not describe the data. The best *s*-wave fit with a free gap parameter $\varepsilon$ results in an unreasonable $\varepsilon=1.94$. The best fit is obtained for a *T*-linear variation of $\lambda_L(T)$, consistent with the clean *d*-wave picture. [9]

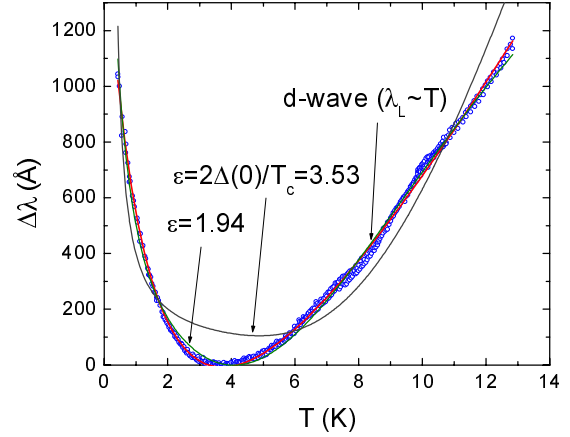

Figure 2. $\Delta\lambda(T)$ in NCCO single crystal

In conclusion, we have measured low temperature penetration depth in electron-doped NCCO and PCCO single crystals. In both cases a power law variation is observed. Our results strongly suggest the presence of the nodes in the superconducting gap.

We thank J. R. Cooper and S. M. Anlage for useful discussions.

This work was supported by Science and Technology Center for Superconductivity Grant No. NSF-DMR 91-20000. Work in Maryland is supported by Grant No. NSF-DMR 97-32736